\documentclass[aps,twocolumn,superscriptaddress]{revtex4}

\usepackage{epsfig}

\begin{document}
\vspace*{1cm}

\title{Parton Energy Loss at Twist-Six in Deeply Inelastic e-A Scattering}

\author{Yun Guo\footnote{E-mail:guoy@iopp.ccnu.edu.cn(Yun Guo)}}
\affiliation{Institute of Particle Physics, Huazhong Normal
             University, Wuhan 430079, China}

\author{Ben-Wei Zhang\footnote{E-mail:bwzhang@iopp.ccnu.edu.cn(Ben-Wei
       Zhang)}}
\affiliation{Institute of Particle Physics, Huazhong Normal
             University, Wuhan 430079, China}
\affiliation{Institut f\"ur Theoretische Physik, Universit\"at
             Regensburg, D-93040 Regensburg, Germany}

\author{Enke Wang\footnote{E-mail:wangek@iopp.ccnu.edu.cn(Enke
       Wang)}}

\affiliation{Institute of Particle Physics, Huazhong Normal
             University, Wuhan 430079, China}

\begin{abstract}
\baselineskip=12pt

Within the framework of the generalized factorization in pQCD, we
investigate the multiple parton scattering and induced parton
energy loss at twist-6 in deeply inelastic e-A scattering with the
helicity amplitude approximation. It is shown that twist-6
processes will give rise to additional nuclear size dependence of
the parton energy loss due to LPM interference effect while its
contribution is power suppressed.

\end{abstract}

\maketitle

Jet quenching, or parton energy loss induced by multiple
scattering in high-energy nuclear collisions has been proposed as
a very sensitive probe of the properties of the hot and dense
medium \cite{WG:92,GP:90}, which recently has given a compelling
theoretical explanation for many exciting experimental phenomena
observed at RHIC, such as the suppression of high $p_T$ single
hadron production, and the disappearance of away-side two hadron
correlation \cite{W:03}. To investigate the radiative energy loss,
many theoretical methods have been developed so far
\cite{Baier,zha,Wied,GLV,WW1,GW1,GW2,ZW1}. Recently, based on the
generalized factorization theorem the twist expansion approach has
been proposed to derive parton energy loss in terms of the
modified parton fragmentation functions in nuclei
\cite{WW:02,GW1,GW2}, and provides the first evidence that the
$A^{2/3}$ dependence of the jet energy loss describes very well
the HERMES data in e-A deeply inelastic scattering (DIS)
\cite{WW:02}. This twist expansion approach has also been applied
to discussing nuclear enhanced effects in the Drell-Yan process
\cite{Fries1} , heavy quark energy loss induced by gluon radiation
in nuclei \cite{ZWW1} and other processes \cite{ZW2,Fries2}.

In this letter, we will extend the twist expansion approach to
study the parton multiple scattering at twist-6 for e-A DIS in
nuclei. In the framework of generalized factorization of high
twist \cite{QS,LQS}, we will derive the semi-inclusive hadronic
tensor at twist-6 with the approach of helicity amplitude
approximation(HAA) \cite{GW2} and the corresponding parton energy
loss. Also we will show the nuclear size $R_A$ dependence of
parton energy loss at twist-6 due to the
Landau-Pomeranchuk-Migdal(LPM) interference effect\cite{LPM}.

We consider the semi-inclusive process in e-A DIS, $e(L_1) + A(p)
\longrightarrow e(L_2) + h (\ell_h) +X$. The differential cross
section for this process has a form as
\begin{equation}
\label{sec1}
    E_{L_2}E_{\ell_h}\frac{d\sigma_{\rm DIS}^h}{d^3L_2d^3\ell_h}
    =\frac{\alpha^2_{\rm EM}}{2\pi s}\frac{1}{Q^4} L_{\mu\nu}
    E_{\ell_h}\frac{dW^{\mu\nu}}{d^3\ell_h}\, ,
\end{equation}
where $p = [p^+,0,{\bf 0}_\perp]$ is the momentum per nucleon in
the nucleus with the atomic number $A$, $q =L_2-L_1 = [-Q^2/2q^-,
q^-, {\bf 0}_\perp]$ is the momentum transfer, the Bjorken
variable is defined as $x_B=Q^2/2p^+q^-$, $s=(p+L_1)^2$ and
$\alpha_{\rm EM}$ is the electromagnetic (EM) coupling constant.
The leptonic tensor is $L_{\mu\nu}=1/2\, {\rm Tr}(\gamma \cdot L_1
\gamma_{\mu} \gamma \cdot L_2 \gamma_{\nu})$ while the
semi-inclusive hadronic tensor is defined as
\begin{eqnarray}
\label{sec2}
   E_{\ell_h}\frac{dW_{\mu\nu}}{d^3\ell_h}&=&
   \frac{1}{2}\sum_X \langle A|J_\mu(0)|X,h\rangle
   \langle X,h| J_\nu(0)|A\rangle \nonumber \\
   &\times &2\pi \delta^4(q+p-p_X-\ell_h),
\end{eqnarray}
here $\sum_X$ runs over all possible final states and
$J_\mu=\sum_q e_q \bar{\psi}_q \gamma_\mu\psi_q$ is the hadronic
EM current.

In the parton model with collinear factorization approximation,
the leading-twist contribution at the lowest order of single
scattering can be factorized as,
\begin{eqnarray}
\label{W1} \frac{dW^S_{\mu\nu}}{dz_h}& =& \sum_q  \int dx
f_q^A(x,\mu_I^2)\nonumber \\
&\times& H^{(0)}_{\mu\nu}(x,p,q)
D_{q\rightarrow h}(z_h,\mu^2)\, \nonumber \\
H^{(0)}_{\mu\nu}(x,p,q)& =& \frac{e_q^2}{2}\, {\rm Tr}(\gamma
\cdot p \gamma_{\mu} \gamma \cdot(q+xp) \gamma_{\nu})
\, \nonumber \\
&\times&\frac{2\pi}{2p\cdot q} \delta(x-x_B) \, ,
\end{eqnarray}
where the momentum fraction carried by the hadron is defined as
$z_h=\ell_h^-/q^-$. $H^{(0)}_{\mu\nu}$ is the hard partonic
tensor, $\mu_I^2$ and $\mu^2$ are the factorization scales for the
initial quark distributions $f_q^A(x,\mu_I^2)$ in a nucleus and
the fragmentation functions $D_{q\rightarrow h}(z_h,\mu^2)$,
respectively. Including all leading log radiative corrections, the
renormalized quark fragmentation function $D_{q\rightarrow
h}(z_h,\mu^2)$ satisfies the
Dokshitzer-Gribov-Lipatov-Altarelli-Parisi (DGLAP) evolution
equations \cite{g}.

\begin{figure}
\centerline{\psfig{file=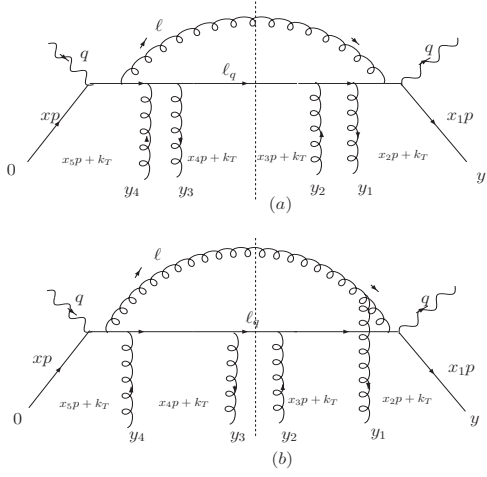,width=200pt,height=180pt}}
\caption{\small Sample diagrams for rescattering with gluon
    radiation in deeply inelastic e-A scattering. $x$, $x_1$, $x_2$,
    $x_3$, $x_4$ and $x_5$ are the momentum fractions carried by initial
    parton. $0$, $y$, $y_1$, $y_2$, $y_3$ and $y_4$ represent the
    positions of the rescattering.}
\label{fig1}
\end{figure}

In the nuclei the propagating quark in DIS will experience
additional scatterings with other partons from the nucleus. The
rescatterings may induce additional gluon radiation which will
cause the energy loss of the leading quark. Such phenomena are
associated with the high twist effect. Previous results
\cite{WW:02,GW1,GW2,ZW1,ZWW1,ZW2} have discussed the double
scattering which gives twist-4 contribution. Here we would like to
go beyond double scattering and extend the study to investigate
the twist-6 (triple scattering) processes as illustrated in
Fig.~\ref{fig1}, and their contribution to the parton energy loss.

According to the generalized factorization theorem of high twist
processes \cite{Fries2,QS,LQS}, the leading twist-6 (triple
scattering) contribution with nuclear medium effect to the
semi-inclusive hadronic tensor can be formulated as
\begin{eqnarray}
\label{W2}
  &&\frac{dW_{\mu\nu}^T}{dz_h} =\sum_q
   \,\int_{z_h}^1\frac{dz}{z}D_{q\rightarrow h}(z_h/z)\nonumber \\
  &&\times
   \int \frac{dy^{-}}{2\pi}\,dy_1^-dy_2^-dy_3^-dy_4^-
   T^{'A}_{qgg}(y^-,y_1^-,y_2^-,y_3^-,y_4^-)\nonumber \\
  &&\times(-\frac{1}{2}g^{\alpha\beta})(-\frac{1}{2}g^{\rho\sigma})
   \Bigl[ \frac{1}{4!}
   \frac{\partial^{2}}{\partial k_{T}^{\alpha} \partial k_{T}^{\beta}}\,
   \frac{\partial^{2}}{\partial k_{T}^{\rho} \partial k_{T}^{\sigma}}\nonumber \\
  &&\quad\overline{H}_{\mu\nu}^T(y^-,y_1^-,y_2^-,y_3^-,y_4^-,k_T,p,q,z)
   \Bigr]_{k_T=0}\,,
\end{eqnarray}
where $k_T$ is the initial gluon's intrinsic transverse momentum
and $T^{'A}_{qgg}$ is related to the parton correlation function
of one quark and two gluons in a nucleus,
\begin{eqnarray}
\label{TA}
  &&T^{'A}_{qgg}(y^-,y_1^-,y_2^-,y_3^-,y_4^-)=\frac{1}{2}\,
     \langle A | \bar{\psi}_q(0)\,
     \gamma^+\,\nonumber\\
  &&~ F_{\tau}^{\ +}(y_{4}^{-})\,
    F_{\upsilon}^{\ +}(y_3^{-})
    F^{+\upsilon}(y_{2}^{-})\,
    F^{+\tau}(y_1^{-})\,\psi_q(y^{-})
     | A\rangle\, .
\end{eqnarray}
Because of the collinear approximation, the tensor structure of
the triple scattering is generally the same as in the leading
order single scattering, $\overline{H}_{\mu\nu}^T$ in
Eq.~(\ref{W2}) can be expressed as
\begin{eqnarray}
\label{H}
  &&\overline{H}_{\mu\nu}^T
   (y^-,y_1^-,y_2^-,y_3^-,y_4^-,k_T,p,q,z){=}
   \int dx H_{\mu\nu}^{(0)}(x,p,q)\nonumber\\
  &&\qquad\qquad\times  \overline{H}^T
   (y^-,y_1^-,y_2^-,y_3^-,y_4^-,k_T,x,p,q,z)\, .
\end{eqnarray}

At twist-4 \cite{WW:02,GW1,GW2,ZW1,ZWW1} there are 23
cut-diagrams, while at twist-6 there are 201 cut-diagrams. To
simplify our calculations, we adopt the helicity amplitude
approximation \cite{GW1,GW2} with the soft gluon radiation
$z_g=(1-z)\rightarrow 0$, where $z_g$ is the momentum fraction
carried by the radiated gluon and $z$ is the momentum fraction
carried by the leading quark. With helicity amplitude
approximation, it is similar to the twist-4 processes that the
leading order contribution comes from 49 central-cut diagrams for
the twist-6 processes.

\begin{figure*}
\centerline{\psfig{file=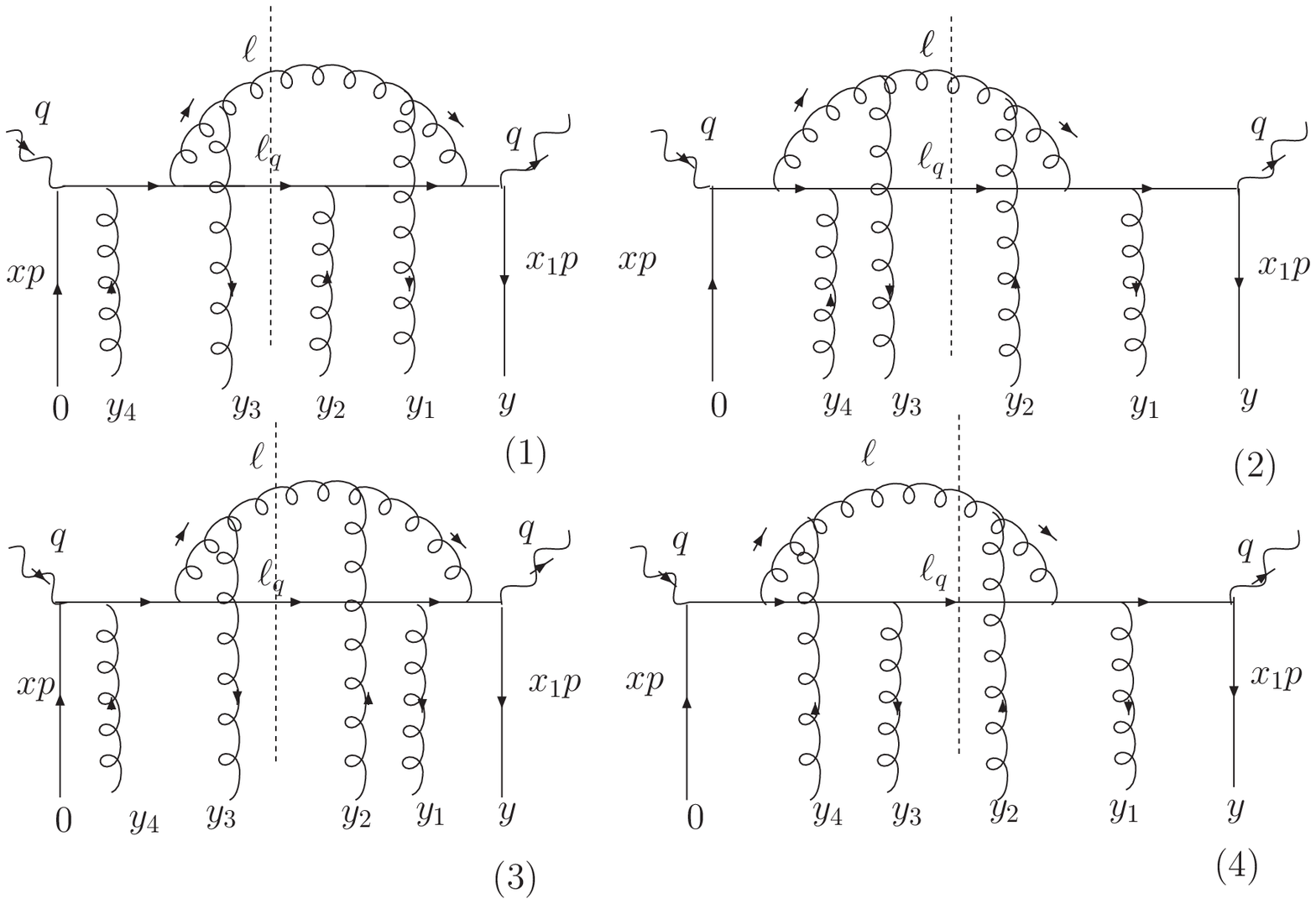,width=300pt,height=140pt}}
\end{figure*}
\begin{figure*}
\centerline{\psfig{file=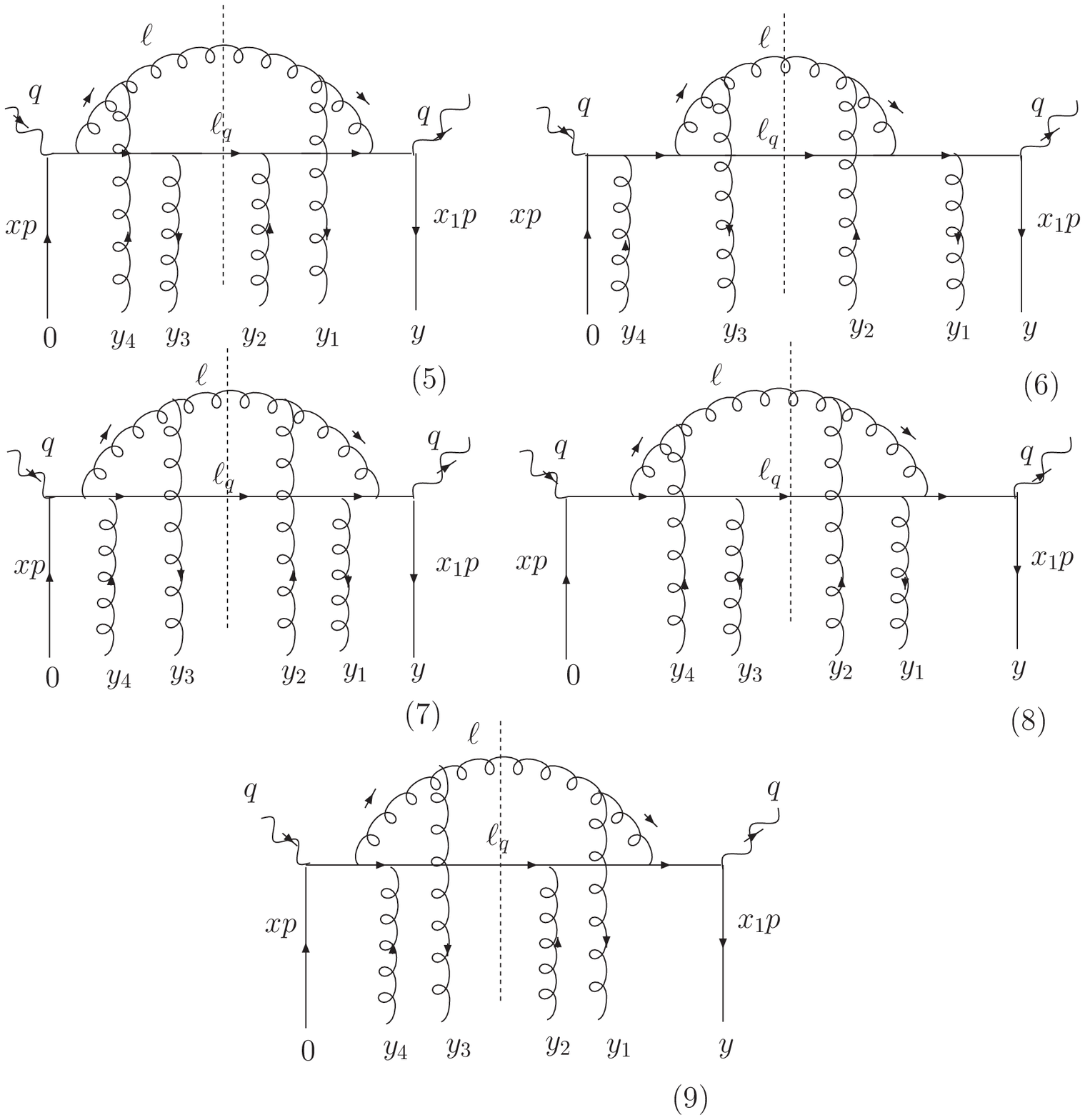,width=300pt,height=210pt}}
\caption{\small Diagrams of the twist-6 processes with gluon
    radiation and non-zero contribution to the hadronic tensor in
    deeply inelastic e-A scattering.}
\label{fig2}
\end{figure*}

For the contributions of the diagrams in Fig.~\ref{fig1}, the
results can be expressed as
\begin{eqnarray}
     \overline{H}_{(a)}^{T}&\propto&\frac{1}{\ell_T^{2}}(1-e^{ix_Lp^+y_4^-})\nonumber \\
  &\times&e^{i(x_B+x_L)p^+y^- + ix_Dp^+(y_1^- - y_2^- + y_3^- -
     y_4^-)}\nonumber \\
  &\times& (1-e^{ix_Lp^+(y^- - y_1^-)})\; ,
\label{Ha}\\
     \overline{H}_{(b)}^{T}&\propto&\frac{ \vec{\ell_T} \cdot
     (\vec{\ell_T}-\vec{k_T}) }
     {\ell_T^2(\vec{\ell_T}-\vec{k_T})^2}(1-e^{ix_Lp^+y_4^-})\nonumber \\
  &\times& e^{i(x_B+x_L)p^+y^- + ix_Dp^+(y_1^- - y_2^- + y_3^- -
     y_4^-)} \nonumber \\
  &\times& (e^{-ix_Lp^+(y^- - y_1^-)}-e^{ix_Dp^+(y^- -
     y_1^-)/(1-z)}) \; .
\label{Hb}
\end{eqnarray}

It is similar to the results of Eqs. (\ref{Ha}) and (\ref{Hb}),
one can check that the contributions from 40 diagrams among the 49
central-cut diagrams have the form
\begin{eqnarray}
\label{Hab}
    \overline{H}^T\propto\frac{ \vec{\ell_T} \cdot
    (\vec{\ell_T}-\eta\vec{k_T}) }
    {\ell_T^2(\vec{\ell_T}-\eta\vec{k_T})^2} e^{iXp^+Y^-},
\end{eqnarray}
where we introduced a symbol $\eta=0,1$, $X$ and $Y^{-}$ represent
the longitudinal momentum fraction and the spatial coordinates,
respectively. In Eq.~(\ref{Ha}), the form of
$\overline{H}^{T}_{(a)}$ corresponds to the case for $\eta=0$ and
$\overline{H}^{T}_{(b)}$ in Eq.~(\ref{Hb}) is the case for
$\eta=1$. Other 38 diagrams have the similar results. According to
the above factorization formula, we can prove that the fourth
derivative of the above expression with respect to $k_{T}$
vanishes at $k_{T}$=0 when we keep only the leading terms as
$\ell_T \rightarrow 0$. Therefore, the contributions from these 40
central-cut diagrams to the hadronic tensor will vanish. As a
result, only 9 central-cut diagrams as shown in Fig.~\ref{fig2}
give non-zero contribution to the hadronic tensor with helicity
amplitude approximation.

For the central-cut diagrams in Fig.~\ref{fig2}, the contributions
have the following form
\begin{eqnarray}
\label{Hi}
     \overline{H}_i^T&=& \int
     \frac{d\ell_T^2}{(\vec{\ell}_T-\vec{k}_T)^{2}}\,
     \frac{\alpha_s}{2\pi}\,\frac{1+z^2}{1-z}16\pi^2\alpha_s^2 \nonumber \\
  &\times&e^{i(x+x_L)p^+y^- + ix_Dp^+(y_1^- - y_2^- + y_3^- -
     y_4^-)}\theta(-y_4^-)\nonumber \\
  &\times&\theta(y_4^- - y_3^-)\theta(y_1^- - y_2^-)\theta(y^- -
     y_1^-)\overline{I}_i\, .
\end{eqnarray}
For each diagram in Fig.~\ref{fig2}, $\overline{I}_i$ can be
expressed as
\begin{eqnarray}
     \overline{I}_1&=&(e^{-i(x_D^0-x_D)p^+(y_4^-
     -y^-_3)-ix_Lp^+y^-_3} \nonumber \\
  &-&e^{i(1-z/(1-z))x_Dp^+(y_4^-
     -y^-_3)-ix_Lp^+y^-_4}) \nonumber \\
  &\times &(e^{-ix_Lp^+(y^- - y_1^-)}-e^{ix_Dp^+(y^- -y_1^-)/(1-z)})\; ,
\label{I1}\\
     \overline{I}_2&=&e^{-i(1-z/(1-z))x_Dp^+(y_3^- -
     y_4^-)}\nonumber \\
  &\times &(e^{-ix_Lp^+
     y_4^-}-e^{ix_Dp^+y_4^-/(1-z)}) \nonumber \\
  &\times & (e^{i(x_D^0-x_D)p^+(y_1^- -y^-_2)-ix_Lp^+(y^- -y^-_2)}\nonumber \\
  &-&e^{-i(1-z/(1-z))x_Dp^+(y_1^- -y^-_2)-ix_Lp^+(y^- -y^-_1)})\; ,
\label{I2}\\
     \overline{I}_3&=&e^{-i(1-z/(1-z))x_Dp^+(y_1^- -
     y_2^-)}\nonumber \\
  &\times &(e^{-ix_Lp^+(y^- -
     y_1^-)}-e^{ix_Dp^+(y^- -y_1^-)/(1-z)}) \nonumber \\
  &\times & (e^{-i(x_D^0-x_D)p^+(y_4^- -y^-_3)-ix_Lp^+y^-_3}\nonumber \\
  &-&e^{i(1-z/(1-z))x_Dp^+(y_4^- -y^-_3)-ix_Lp^+y^-_4})\; ,
\label{I3}\\
     \overline{I}_4&=&(e^{-ix_Lp^+y_4^-}-e^{ix_Dp^+y_4^-/(1-z)})\nonumber \\
  &\times &(e^{i(x_D^0-x_D)p^+(y_1^- -y^-_2)-ix_Lp^+(y^- -y^-_2)}\nonumber\\
  &-&e^{-i(1-z/(1-z))x_Dp^+(y_1^- -y^-_2)-ix_Lp^+(y^- -y^-_1)}) \; ,
\label{I4}\\
     \overline{I}_5&=&(e^{-ix_Lp^+y_4^-}-e^{ix_Dp^+y_4^-/(1-z)})\nonumber \\
  &\times&(e^{-ix_Lp^+(y^- - y_1^-)}-e^{ix_Dp^+(y^- -y_1^-)/(1-z)}) \; ,
\label{I5}\\
     \overline{I}_6&=&(e^{-i(x_D^0-x_D)p^+(y_4^- -y^-_3)-ix_Lp^+y^-_3} \nonumber \\
  &-&e^{i(1-z/(1-z))x_Dp^+(y_4^- -y^-_3)-ix_Lp^+y^-_4}) \nonumber \\
  &\times &(e^{i(x_D^0-x_D)p^+(y_1^- -y^-_2)-ix_Lp^+(y^- - y^-_2)}\nonumber \\
  &-&e^{-i(1-z/(1-z))x_Dp^+(y_1^- -y^-_2)-ix_Lp^+(y^- -y^-_1)}) \; ,
\label{I6}\\
    \overline{I}_7&=&e^{-i(1-z/(1-z))x_Dp^+(y_1^- - y_2^- + y_3^- - y_4^-)}\nonumber \\
  &\times &(e^{-ix_Lp^+y_4^-}-e^{ix_Dp^+y_4^-/(1-z)}) \nonumber \\
  &\times&(e^{-ix_Lp^+(y^- - y_1^-)}-e^{ix_Dp^+(y^- -y_1^-)/(1-z)})\; ,
\label{I7}\\
    \overline{I}_8&=&e^{-i(1-z/(1-z))x_Dp^+(y_1^- - y_2^-)}\nonumber \\
  &\times &(e^{-ix_Lp^+y_4^-}-e^{ix_Dp^+y_4^-/(1-z)}) \nonumber \\
  &\times&(e^{-ix_Lp^+(y^- - y_1^-)}-e^{ix_Dp^+(y^- -y_1^-)/(1-z)})\; ,
\label{I8}\\
    \overline{I}_9&=&e^{-i(1-z/(1-z))x_Dp^+(y_3^- - y_4^-)} \nonumber \\
  &\times &(e^{-ix_Lp^+y_4^-}-e^{ix_Dp^+y_4^-/(1-z)})\nonumber \\
  &\times&(e^{-ix_Lp^+(y^- - y_1^-)}-e^{ix_Dp^+(y^- -y_1^-)/(1-z)})\; .
\label{I9}
\end{eqnarray}
The variables $x_L$, $x_D^0$ and $x_D$ in the above expressions
are defined as
\begin{eqnarray}
   x_L &=& \frac{\ell_T^2}{2p^+q^-z(1-z)}\, ,\\
\label{xL}
   x_D^0 &=& \frac{k_T^2}{2p^+q^-}\, , \\
\label{xD0}
   x_D &=& \frac{k_T^2-2\vec{k}_T\cdot \vec{\ell}_T}{2p^+q^-z}\, ,
\label{xD}
\end{eqnarray}
where $\ell_T$ is the transverse momentum of the radiated gluon,
and $z=\ell_q^-/q^-$ is the momentum fraction carried by the final
quark. Four terms of each expression in Eqs. (\ref{I1}-\ref{I9})
represent the contributions from different processes and their
interferences. In the collinear limit as $\ell_T \rightarrow 0$
and $k_T \rightarrow 0$, there will be a clear cancelation of
these contributions due to the LPM effect. We will see later that
the LPM effect will give an additional nuclear size dependence to
the parton energy loss.

Adding up all contributions together, we can derive the
semi-inclusive hadronic tensor for triple scattering processes at
twist-6,
\begin{eqnarray}
\label{WT}
    \frac{W_{\mu\nu}^{T,q}}{dz_h} &=&\sum_q \,\int dx
    H^{(0)}_{\mu\nu}(x,p,q) \frac{\alpha_s}{2\pi}
    \int_{z_h}^1\frac{dz}{z}D_{q\rightarrow h}(z_h/z)\, \nonumber \\
  &\times&\frac{2}{3}\int \frac{d\ell_T^2}{\ell_T^6} 16\pi^2\alpha_s^2
    [\frac{1+z^2}{(1-z)}_{+}T^{A}_{qgg}(x,x_L)\nonumber\\
  &+&\delta(z-1)\Delta T^{A}_{qgg}(x,\ell_T^2)]\, ,
\end{eqnarray}
where the $"+"$ functions have a form
\begin{equation}
\label{+}
     \int_0^1 dz \frac{F(z)}{(1-z)_+} \equiv \int_0^1 dz
     \frac{F(z)-F(1)}{1-z},
\end{equation}
$F(z)$ being any function which is sufficiently smooth at $z=1$.
Here we also include the virtual corrections in Eq. (\ref{WT}),
which is obtained by using the unitarity requirement and ensures
the final result to be infrared safe \cite{GW1}.

The parton correlation function $T^{A}_{qgg}(x,x_L)$ in Eq.
(\ref{WT}) is expressed as
\begin{eqnarray}
\label{TAqgg}
     &&T^{A}_{qgg}(x,x_L)=\int\frac{dy^{-}}{2\pi}\,dy_1^-dy_2^-dy_3^-dy_4^-
     e^{i(x+x_L)p^+y^-}\nonumber \\
   &&\quad\times
     \theta(y_1^- - y_2^-)\theta(y^- - y_1^-)\theta(-y_4^-)\theta(y_4^- - y_3^-)\nonumber\\
   &&\quad\times
     I(x_L)T^{'A}_{qgg}(y^-,y_1^-,y_2^-,y_3^-,y_4^-)\, ,
\end{eqnarray}
where function $I(x_L)$ is defined as
\begin{eqnarray}
\label{I}
     I(x_L)&=&\sum\limits_{i=1,2}\sum\limits_{j=3,4}C_{ij}(1-e^{-ix_Lp^+y^-_j})\nonumber\\
  &\times&(1-e^{-ix_Lp^+(y^- -y^-_i)}).
\end{eqnarray}
Here $C_{ij}$ are associated with the color factors of each
cut-diagram, and we have $C_{14}=\frac{35}{96}$,
$C_{24}=C_{13}=-\frac{9}{64}$ and $C_{23}=\frac{1}{12}$. In
principle, the parton correlation function $T^{A}_{qgg}(x,x_L)$ is
not calculable and can only be measured in the experiments.
However, under some assumptions, one can relate this parton
correlation function to parton distributions and estimate its
value \cite{GW2,Fries2,LQS}. The definition of $\Delta
T^{A}_{qgg}(x,\ell_T^2)$  in Eq. (\ref{WT}) is
\begin{eqnarray}
\label{DT}
     \Delta T^{A}_{qgg}(x,\ell_T^2) &= & \int_0^1
     dz\frac{1}{1-z}[ 2 T^{A}_{qgg}(x,x_L)|_{z=1} \nonumber\\
  &-&(1+z^2) T^{A}_{qgg}(x,x_L)] \, .
\end{eqnarray}

Considering the high-twist contributions up to twist-6, we can
express the hadronic tensor as
\begin{eqnarray}
\label{W3}
     \frac{dW_{\mu\nu}}{dz_h}&=&\sum_q \int dx
     \widetilde{f}_q^A(x,\mu_I^2) H^{(0)}_{\mu\nu}(x,p,q)\nonumber\\
  &\times & \widetilde{D}_{q\rightarrow h}(z_h,\mu^2),
\end{eqnarray}
where the quark distribution function
$\widetilde{f}_q^A(x,\mu_I^2)$ include also the twist-6
contributions. Up to twist-6 the modified quark fragmentation
function $\widetilde{D}_{q\rightarrow h}(z_h,\mu^2)$ has the form
\begin{eqnarray}
\label{MD}
     &&\widetilde{D}_{q\rightarrow h}(z_h,\mu^2)=D_{q\rightarrow h}(z_h,\mu^2)
     +D^{(twist-4)}_{q\rightarrow h}(z_h,\mu^2)\nonumber\\
  &&\qquad\qquad\qquad\quad +D^{(twist-6)}_{q\rightarrow h}(z_h,\mu^2)\nonumber\\
  &&\quad =D_{q\rightarrow h}(z_h,\mu^2)
     +D^{(twist-4)}_{q\rightarrow h}(z_h,\mu^2)
     +\int_0^{\mu^2} \frac{d\ell_T^2}{\ell_T^2}\nonumber\\
  &&\quad\times\frac{\alpha_s}{2\pi}\int_{z_h}^1\frac{dz}{z}
     [\Delta\gamma^{(twist-6)}_{q\rightarrow qg}(z,x,x_L,\ell_T^2)
     D_{q\rightarrow h}(z_h/z)\nonumber\\
  &&\quad +\Delta\gamma^{(twist-6)}_{q\rightarrow gq}(z,x,x_L,\ell_T^2)
     D_{g\rightarrow h}(z_h/z)]\, ,
\end{eqnarray}
where quark fragmentation function $D^{(twist-4)}_{q\rightarrow
h}(z_h,\mu^2)$ together with corresponding modified splitting
functions $\Delta\gamma^{(twist-4)}_{q\rightarrow
qg}(z,x,x_L,\ell_T^2)$ and $\Delta\gamma^{(twist-4)}_{q\rightarrow
gq}(z,x,x_L,\ell_T^2)$ at twist-4 have been derived in Ref.
\cite{GW1,GW2}, the third term $D^{(twist-6)}_{q\rightarrow
h}(z_h,\mu^2)$ in Eq.~(\ref{MD}) is the additional corrections to
quark fragmentation function at twist-6. In Eq.~(\ref{MD}) the
modified splitting functions at twist-6 are
\begin{eqnarray}
     &&\Delta\gamma^{(twist-6)}_{q\rightarrow qg}(z,x,x_L,\ell_T^2)
     =\frac{2(4\pi\alpha_s)^2}
     {3(\ell_T^2+\langle k_T^2\rangle)^2\widetilde{f}_q^A(x,\mu_I^2)}\nonumber \\
  &&\quad\times [\frac{1{+}z^2}{(1{-}z)_+}T^{A}_{qgg}(x,x_L)
     {+}\delta(1{-}z)\Delta T^{A}_{qgg}(x,\ell_T^2)]\, ,
\label{gammaqqg}\\
  &&\Delta\gamma^{(twist-6)}_{q\rightarrow gq}(z,x,x_L,\ell_T^2)\nonumber\\
  &&\quad=\Delta\gamma^{(twist-6)}_{q\rightarrow qg}(1-z,x,x_L,\ell_T^2)\, .
\label{gammaqgq}
\end{eqnarray}
Here, we replace $1/\ell_T^6$ in Eq.(\ref{WT}) with
$1/\ell_T^2(\ell_T^2+\langle k_T^2\rangle)^2$ in order to restore
the collinear structure of the scattering amplitude \cite{GW2}.

In Eq.(\ref{WT}) for the hadronic tensor at twist-6, there is an
additional factor $1/\ell_T^2$ as compared to the case at twist-4
\cite{GW1,GW2}. For large $\ell_T^2$, LPM effect becomes
unimportant and one can neglect the interference contribution. In
these processes, large final transverse momentum $\ell_T^2\sim
Q^2$ will lead to a $1/Q^2$ suppression as compared to the case at
twist-4 \cite{QS,LQS}. On the other hand, when $\ell_T^2$ takes an
intermediate value, we can not ignore LPM effect. One can check in
Eq.(\ref{I}), if $\ell_T\rightarrow 0$, there will be a clear
cancelation by the interferences. The LPM interference effect
restricts the radiated gluon to have a minimum transverse momentum
$\ell_T^2\sim Q^2/A^{1/3}$ \cite{GW2}. Therefore, the additional
factor $1/\ell_T^2$ in Eq.~(\ref{WT}) as compared to the case at
twist-4 has a suppression of $R_A/Q^2$ for intermediate $Q^2$ at
twist-6 due to the LPM interference effect, here $R_A=1.12A^{1/3}$
is the radius of the nucleus.

As shown in Ref.~\cite{GW2,WW:02}, the fractional energy loss of
the leading quark can be given by the fractional energy carried
away by the induced gluon, the fractional energy loss of the
leading quark at twist-4 can be expressed as
\begin{eqnarray}
      &&\langle\Delta z_g\rangle _{twist-4}
      =\int_0^{\mu^2}\frac{d\ell_T^2}{\ell_T^2} \int_0^1 dz
      \frac{\alpha_s}{2\pi} z\,\nonumber \\
  &&\qquad\times \Delta\gamma^{(twist-4)}_{q\rightarrow
      gq}(z,x_B,x_L,\ell_T^2)\nonumber\\
  &&\quad =\int_0^{\mu^2}\frac{d\ell_T^2}{\ell_T^2} \int_0^1 dz
      \frac{1{+}(1{-}z)^2}{\ell_T^2{+}\langle k_T^2\rangle}
      \frac{\alpha_s^2 T_{qg}^A (x_B,x_L)}{\widetilde{f}_q^A(x_B)}\, ,
 \label{E4}
\end{eqnarray}
here $\widetilde{f}_q^A(x_B)$ is the quark distribution function
up to the twist-4. For the twist-4 processes, the parton
correlation function $T^{A}_{qg}(x,x_L)$ contributes a $R_A$
dependence\cite{GW2,WW:02}. After taking into account
$1/\ell_T^2\sim R_A/Q^2$ resulting from the LPM effect, the
nuclear correction to the fragmentation function and the parton
energy loss due to double parton scattering (twist-4) processes
will then be in the order of $\alpha_s^2 R_A/\ell_T^2\sim
\alpha_s^2 R_A^2/Q^2$. As shown in Ref.~\cite{GW2,WW:02}, the
order of the fractional energy loss for leading quark has a form
\begin{equation}
    \langle\Delta z_g\rangle _{twist-4}\sim \alpha_s^2\frac{x_B}{x_A^2 Q^2} \, ,
\label{dE4}
\end{equation}
where $x_A=1/m_N R_A$ with $m_N$ is the nucleon's mass.

At twist-6 it is the same as in the case of the twist-4 the
fractional energy loss of the leading quark is given by the
fractional energy carried away by the induced gluon, then the
fractional energy loss of the leading quark has a form
\begin{eqnarray}
       &&\langle\Delta z_g\rangle _{twist-6}
       =\int_0^{\mu^2}\frac{d\ell_T^2}{\ell_T^2} \int_0^1 dz
       \frac{\alpha_s}{2\pi} z\nonumber \\
   &&\qquad\times \Delta\gamma^{(twist-6)}_{q\rightarrow
      gq}(z,x_B,x_L,\ell_T^2)\nonumber\\
   &&\quad=\int_0^{\mu^2}\frac{d\ell_T^2}{\ell_T^2} \int_0^1 dz
      \frac{1+(1-z)^2}{(\ell_T^2+\langle
      k_T^2\rangle)^2}\nonumber\\
   &&\qquad\times
      \frac{16\pi\alpha_s^3 T_{qgg}^A (x_B,x_L)}{3\widetilde{f}_q^A(x_B)}\, ,
 \label{E6}
\end{eqnarray}
here quark distribution function $\widetilde{f}_q^A(x_B)$ contains
the contribution of the twist-6 processes. The parton correlation
function $T^{A}_{qgg}(x,x_L)$ at twist-6 will contribute $R_A^2$
dependence because $T^{A}_{qgg}(x,x_L)$ has an additional $R_A$
dependence due to the extra gluon pair as compared to the case at
twist-4. Here we adopt also the assumption \cite{GW2,Fries2,LQS}
that the positions of the two extra gluon field strengths are
confined within one nucleon. As has been shown $1/\ell_T^2\sim
R_A/Q^2$ due to the LPM effect, combine $R_A^2$ dependence of the
parton correlation function $T^{A}_{qgg}(x,x_L)$, the nuclear
correction to the fragmentation function and the parton energy
loss due to triple parton scattering (twist-6) processes will then
be in the order of $\alpha_s^3 R_A^2/\ell_T^4\sim
\alpha_s^3R_A^4/Q^4$. The order of the fractional energy loss for
leading quark can be expressed as
\begin{equation}
\label{dE6}
    \langle\Delta z_g\rangle _{twist-6}\sim\alpha_s^3\frac{x_B^2}{x_A^4 Q^4} \, .
\end{equation}
One can see $\langle\Delta z_g\rangle _{twist-6}$ has $\alpha_s
R_A^2/Q^2$ suppression as compared to $\langle\Delta z_g\rangle
_{twist-4}$ for fixed $x_B$.

The twist-6 calculation indicates that the twist expansion by
taking into account the multiple scattering in nuclear medium is
actually the expansion with the parameter $\alpha_s R_A^2/Q^2$ due
to LPM effect. Whereas the LPM effect can be neglected, the
expanding parameter will be $\alpha_s R_A/Q^2$, so that we can see
the LPM effect plays an important role to give an additional
nuclear size dependence for the parton energy loss. Furthermore,
we can see, as $Q^2$ is large and $R_A$ is not very large, the
parameter $\alpha_s R_A^2/Q^2$ gives a small value and it is
reasonable to consider only the double scattering processes
because all high twist contribution will be suppressed by the
power of $\alpha_s R_A^2/Q^2$. However, for the intermediate $Q^2$
and large $R_A$ the high twist processes will give considerable
contribution and the resummation of all high twist contribution
will be needed. This resummation would be similar to resuming all
power corrections to the nuclear structure functions in e-A DIS
\cite{qiu}, except that in our case the calculations may be
complicated by the additional induced gluon radiation. We expect
the resummation of all high twist contributions to parton energy
loss may adopt a similar reaction operator formalism as derived by
GLV to compute the radiative energy loss to all orders in opacity
in a hot QCD medium \cite{GLV}, and this work is in progress.

In summary, we have calculated the semi-inclusive hadronic tensor
and the corresponding parton energy loss at twist-6 in e-A DIS
with helicity amplitude approximation by utilizing the twist
expansion approach. The energy loss induced by gluon radiation at
triple scattering (twist-6) processes has a $\alpha_s R_A^2/Q^2$
suppression due to LPM interference effect as compared to the
contribution of double scattering (twist-4) processes. The
expanding parameter of high twist-expansion is discussed and we
find the resummation of all high twist processes should be needed
for intermediate $Q^2$ and heavy nucleus with large nucleus
radius.

\section*{Acknowledgments}

The authors are very grateful to Xin-Nian Wang for helpful
discussions. This work was supported by National Natural Science
Foundation of China under project Nos. 10405011, 10475031,
10440420018, by MOE of China under projects NCET-04-0744,
SRFDP-20040511005, CFKSTIP-704035 and by Alexander von Humboldt
Foundation.

\end{document}